# Instantons or Monopoles? Dyons

V. Bornyakov

*Institute for High Energy Physics*
*RU-142284 Protvino, Russia*

e-mail: bornyakov@mx.ihep.su

G. Schierholz

*Deutsches Elektronen-Synchrotron DESY*
*D-22603 Hamburg, Germany*

and

*Gruppe Theorie der Elementarteilchen, Höchstleistungsrechenzentrum HLRZ*
*c/o Forschungszentrum Jülich, 52425 Jülich, Germany*

e-mail: gsch@x4u2.desy.de

## ABSTRACT

Non-abelian gauge theories can be cast into abelian gauge theories with monopoles. We ask what becomes of the instantons after abelian projection. Instantons are found to consist of closed dyon loops. It is shown that the electric charge of the dyons is quantized. The implication of this result for the dynamics of the Yang-Mills vacuum is briefly discussed.

## 1. Introduction

There are two types of topological excitations in the Yang-Mills vacuum, instantons and monopoles. The discovery of instantons [1] has led to the solution of the $U_A(1)$ problem [2,3,4], while there is evidence [5,6] for the idea [7] that color confinement is caused by monopole condensation. Both of the two excitations seem to be able to explain chiral symmetry breaking [8,9]. Instantons and monopoles interact coherently in the $\theta$ vacuum [10], and this might carry the solution to the strong CP problem [11].

Attempts of deriving an effective theory of the QCD vacuum were either based on instantons [8] or on monopoles [13], and the opinions of the various schools on what are the fundamental variables were largely divided. There is evidence that instantons and monopoles are manifestations of one and the same excitation: dyons [14]. This will perhaps unite the various schools. Certainly it will shed some new light on the dynamics of the QCD vacuum.

To begin with, we shall collect some theoretical arguments which suggest that instantons are made of dyons, and then turn to a quantitative analysis of the problem.

Instantons give rise to an integer valued topological charge

$$Q = \frac{1}{16\pi^2} \int d^4x \, \text{Tr} \, F_{\mu\nu} \widetilde{F}_{\mu\nu} \in \mathbb{Z}, \, \widetilde{F}_{\mu\nu} = \frac{1}{2} \epsilon_{\mu\nu\rho\sigma} F_{\rho\sigma}. \tag{1}$$

A classical (anti-)instanton corresponds to a field configuration of minimal action with $Q = \pm 1$ which is (anti-)self-dual, i.e.

$$F_{\mu\nu} = \pm \widetilde{F}_{\mu\nu}, \tag{2}$$

with + sign for the instanton and − sign for the anti-instanton. In the following we shall restrict ourselves to gauge group $SU(2)$. In this case the action for a classical (anti-)instanton is

$$\beta^{-1} S \equiv \frac{1}{8} \int d^4x \, \text{Tr} \, F_{\mu\nu} F_{\mu\nu} = 2\pi^2, \, \beta = \frac{4}{g^2}. \tag{3}$$

Monopoles arise by fixing the gauge so that only the gauge degrees of freedom of the Cartan subgroup $U(1) \subset SU(2)$ are kept dynamical. The gluonic degrees of freedom of the theory may then be mapped onto 'photons', color electric charges and color magnetic charges, i.e. monopoles [10], by what is called abelian projection. The idea is that for a suitably chosen gauge the long-distance physics is essentially described by the abelian degrees of freedom, which is called abelian dominance. Such a gauge is the so-called maximally abelian gauge [a]

$$D_\mu^3 A_\mu^\pm = 0, \tag{4}$$

where $A_\mu^\pm = A_\mu^1 \pm A_\mu^2$. We shall consider the maximally abelian gauge throughout this paper. An essential feature of this gauge is that it is manifestly renormalizable [15].

The abelian vector potentials $a_\mu^i, i = 1, 2$, refered to as 'photons', are taken to be the diagonal components of the (original) $SU(2)$ gauge potentials after gauge fixing. In the following we drop the index and write $a_\mu = a_\mu^1 = -a_\mu^2$. We define an abelian field strength

$$f_{\mu\nu} = \partial_\mu a_\nu - \partial_\nu a_\mu, \tag{5}$$

which leads to the definition of magnetic currents

$$K_\mu = \frac{1}{4\pi} \partial_\nu \widetilde{f}_{\mu\nu}, \, \widetilde{f}_{\mu\nu} = \frac{1}{2} \epsilon_{\mu\nu\rho\sigma} f_{\rho\sigma}. \tag{6}$$

---

[a]The notation refers actually to the lattice version of the gauge [5].

It is easy to see that the magnetic currents are conserved. In this paper we shall consider periodic volumes. In periodic volumes the magnetic currents form closed loops [16]. We recall that integration of the current density over a three-dimensional region $\Omega$ yields

$$m(\Omega) = \int_\Omega \mathrm{d}^3\sigma_\mu K_\mu = \frac{1}{4\pi} \int_{\partial\Omega} \mathrm{d}^2\sigma_{\mu\nu} \widetilde{f}_{\mu\nu}. \tag{7}$$

Choosing the time direction to be $\mu = 4$, and taking $\Omega$ at constant time, we obtain

$$m(\Omega) = \frac{1}{4\pi} \int_{\partial\Omega} \mathrm{d}^2\sigma_j b_j, \ b_j = \frac{1}{2}\epsilon_{jkl} f_{kl}, \tag{8}$$

which is the magnetic flux through $\partial\Omega$, so that $m(\Omega)$ counts the magnetic charge inside $\Omega$. The magnetic charges obey the Dirac quantization condition $m = 0, \pm\frac{1}{2}, \pm 1, \ldots$. In our notation quarks have color electric charge 1. An elementary monopole has charge $m = \frac{1}{2}$, an anti-monopole $m = -\frac{1}{2}$.

Similarly, one can define electric currents

$$J_\mu = \frac{1}{4\pi} \partial_\nu f_{\mu\nu}. \tag{9}$$

Like the magnetic currents, they are also conserved. If the picture of abelian dominance is correct, a non-vanishing topological charge $Q$ should translate into a non-vanishing abelian charge

$$q = \frac{1}{8\pi^2} \int \mathrm{d}^4 x f_{\mu\nu} \widetilde{f}_{\mu\nu}. \tag{10}$$

The abelian charge can be expressed in terms of the magnetic and electric current as follows:

$$q = 2 \int \mathrm{d}^4 x \mathrm{d}^4 y J_\mu(x) V(x-y) K_\mu(y), \tag{11}$$

where $V(x)$ is the abelian Coulomb potential. This implies that an instanton is always accompanied by a monopole loop plus a loop of electric currents. In other words, instantons are a source for both magnetic and electric charges.

The first part of this observation is not new. It is known to us already since a long time [17] that instantons contain monopole loops. Recently this idea has received further support by work of several other authors [18,19,20] [b].

The color electric charge of the monopole is given by [c]

$$e(\Omega) = \frac{1}{4\pi} \int_\Omega \mathrm{d}^3\sigma_\mu \partial_\nu f_{\mu\nu}. \tag{12}$$

---

[b]The analytical result of Chernodub and Gubarev [18] is probably irrelevant for the quantum (equilibrium) theory as the gauge fixing functional in that calculation is divergent (at least in $\mathbb{R}^4$).
[c]Note the change in normalization compared to our first report [14].

Taking $\Omega$ at constant time as in eq. (8), this gives

$$e(\Omega) = \frac{1}{4\pi} \int_{\partial\Omega} \mathrm{d}^2\sigma_j e_j, \ e_j = f_{4j}. \tag{13}$$

Let us consider an (anti-)instanton now. If the (anti-)self-duality (2) survives the abelian projection, then we would have $f_{\mu\nu} = \pm \tilde{f}_{\mu\nu}$, with + sign for the instanton and − sign for the anti-instanton. If we now insert this into eq. (13), we obtain from eq. (8)

$$e(\Omega) = \pm m(\Omega) = \pm\frac{1}{2}. \tag{14}$$

Thus we would conclude that the monopole loop and the loop of electric currents fall on top of each other, which means that instantons consist of dyon loops, with the electric charge of the dyon being quantized. Note that the Dirac quantization condition itself, as well as its generalization [21], does not make any statement about the electric charge of the dyon.

Actually, for this to happen one does not need to have exact (anti-)self-duality on the abelian level. It is sufficient that $f_{\mu\nu} = \pm \tilde{f}_{\mu\nu}$ holds on the surface $\partial\Omega$, which in fact should well include the charge cloud of the monopole. This is a much weaker assumption.

What does this mean for a monopole moving around in the vacuum? When the monopole hits an (anti-)instanton, the monopole will pick up an electric charge of the appropriate amount (if it was not already charged), and vice versa when it leaves the (anti-)instanton. The net effect is that in the background of a configuration of non-vanishing topological charge the monopoles will, on the average, be electrically charged. This comes not unexpected. It has been shown by Witten [22] that in the $\theta$-vacuum monopoles of magnetic charge $m$ acquire an electric charge of the magnitude $4m\,\theta/2\pi$. A non-vanishing vacuum angle $\theta$ implies a non-vanishing background topological charge, if there is no phase transition in $\theta$.

A quantized electric charge is also a property of the static monopole solutions discussed by Smit and van der Sijs [12] and Simonov [23] in their attempt to model the Yang-Mills vacuum, in contrast to the classical dyon solution [24] which admits arbitrary charges.

## 2. Lattice Investigation

The only assumption we have made in deriving that instantons are made of dyons of quantized charge is that of abelian dominance. This is a reasonable assumption if one is primarily interested in the long-distance properties of the theory, because the off-diagonal terms of the gauge potential, which under gauge transformations transform as charged vector fields [16], are expected to be massive and hence will only propagate over microscopic distances. We shall investigate now, on the lattice, to what extent this assumption is true.

We start from the Wilson action

$$S = \sum_{x,\mu<\nu} (1 - \frac{1}{2}\text{Tr } U_{\mu\nu}(x)). \tag{15}$$

On semi-classical configurations we can use the naive topological charge

$$Q = \frac{1}{32\pi^2} \sum_x \epsilon_{\mu\nu\rho\sigma} \text{Tr } \Pi_{\mu\nu}(x)\Pi_{\rho\sigma}(x), \tag{16}$$

where $\Pi_{\mu\nu}(x) = \frac{1}{2i}(U_{\mu\nu}(x) - U_{\mu\nu}^\dagger(x))$. We fix to the maximally abelian gauge by performing a local gauge transformation on the link matrices, $\tilde{U}_\mu(x) = g(x)U_\mu(x)g^{-1}(x)$, such that the expression

$$R = \sum_{x,\mu} \text{Tr } \sigma_3 \tilde{U}_\mu(x) \sigma_3 \tilde{U}_\mu^\dagger(x) \tag{17}$$

is maximized with respect to the choice of $g$. The abelian projection is done in the standard way [16]. We define the abelian action by

$$s = \sum_{x,\mu<\nu} (1 - \cos\theta_{\mu\nu}(x)), \tag{18}$$

where $\theta_{\mu\nu}(x)$ is the plaquette angle of the abelian gauge potentials. Similarly, we define an abelian 'topological' charge

$$q = \frac{1}{8\pi^2} \sum_x \epsilon_{\mu\nu\rho\sigma} \sin\theta_{\mu\nu}(x) \sin\theta_{\rho\sigma}(x). \tag{19}$$

For the monopole currents we take

$$K_\mu(x) = \frac{1}{4\pi} \epsilon_{\mu\nu\rho\sigma} \partial_\nu \overline{\theta}_{\rho\sigma}(x + \hat{\mu}), \tag{20}$$

where $\overline{\theta}_{\rho\sigma}(x)$ is restricted to $-\pi < \overline{\theta}_{\rho\sigma}(x) \leq \pi$. The magnetic charge is defined by [16]

$$m(f(x,\mu)) = \frac{1}{4\pi} \sum_{p \in \partial f(x,\mu)} \overline{\theta}_p, \tag{21}$$

where $f(x,\mu)$ denotes the elementary cube perpendicular to the $\mu$-direction and $p$ denotes the plaquette. We identify dual and original lattice. The lattice monopole current is conserved in the sense that $\partial_\mu^{(-)} K_\mu(x) = 0$, where $\partial_\mu^{(-)}$ is the lattice backward derivative. That means that $K_\mu$ forms a closed loop on the periodic lattice. The electric current is defined by [d]

$$J_\mu(x) = \frac{1}{4\pi} \partial_\nu^{(-)} \overline{\theta}_{\mu\nu}(x). \tag{22}$$

Using that, we find the electric charge in $f(x, \mu)$ to be

$$e(f(x,\mu)) = J_\mu(x) = \frac{1}{4\pi} \sum_\nu [\overline{\theta}_{\mu\nu}(x) - \overline{\theta}_{\mu\nu}(x - \hat{\nu})]. \tag{23}$$

The position of the plaquettes $\overline{\theta}_{\mu\nu}$ relative to the cube is shown in fig. 1. Unlike the case of the magnetic charge, the electric charge can generally not be expected to be localized in a single elementary cube. The total electric charge is given by

$$e(x_\mu) = \sum_{\{x\}\backslash x_\mu} J_\mu(x), \tag{24}$$

where the sum is over the three-dimensional lattice dual to $x_\mu$. Later on we will also use positive and negative charges

$$e^\pm(f(x,\mu)) = \theta(\pm J_\mu(x)) J_\mu(x), \tag{25}$$
$$e^\pm(x_\mu) = \sum_{\{x\}\backslash x_\mu} \theta(\pm J_\mu(x)) J_\mu(x). \tag{26}$$

This makes sense if the centers of positive and negative charges are sufficiently far apart and the charges are localized. It is evident that $e^+(x_\mu) = -e^-(x_\mu)$ for periodic boundary conditions.

To create an ensemble of instantons, it appears to be natural to cool equilibrium configurations down to semi-classical configurations [25]. To show the anticipated effect, the instantons should be large enough in size though.

We cooled equilibrium configurations on $8^4$, $12^4$ and $16^4$ lattices at $\beta = 2.3, 2.4$ and 2.5 until we reached the plateau. We observed [14] that instantons are surrounded by monopole loops. For the abelian 'topological' charge we found $q = (0.4 - 0.7) Q$. We were, however, not successful in finding large instantons, which allowed us to unambiguously determine the electric charge of the monopole. As we will argue later, there is perhaps a good reason for that, namely that the quantum theory does not support large isolated instantons.

---

[d] One could also use $\sin \theta_{\mu\nu}$ instead of $\overline{\theta}_{\mu\nu}$. We checked that this makes practically no difference on our configurations.

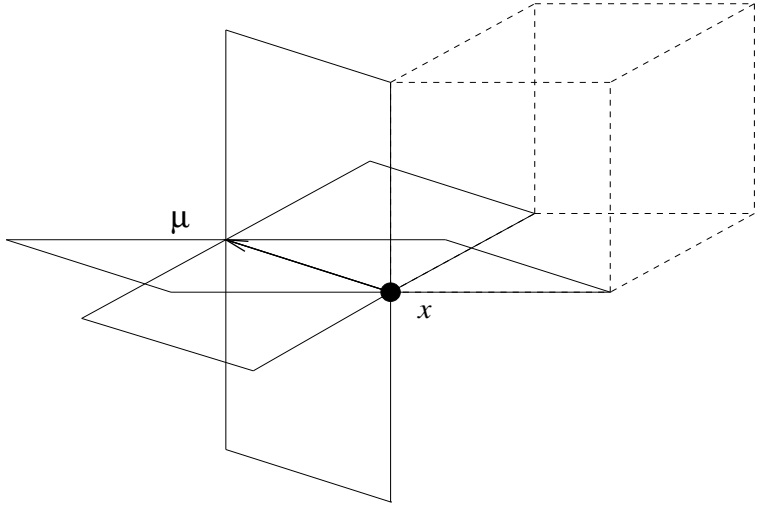

Fig. 1. The current $J_\mu(x)$. The solid lines indicate the position of the plaquettes $\overline{\theta}_{\mu\nu}$ in eq. (23) relative to the cube $f(x,\mu)$ which is marked by dashed lines.

A more direct way of creating large instantons is to start from the periodic (lattice) instanton [26]. This construction is topologically correct, but it is not a minimum of the action. To make the gauge potential smooth at the boundary of the different patches of gauge, we apply several cooling sweeps. For appropriate parameters, such as the core parameter and the box sizes, it was possible to obtain instantons with well separated monopole currents.

We shall now pick two examples. First we will consider a single instanton, and second the case of two well separated instantons, both on a $16^4$ lattice.

Let us first concentrate on the single instanton. A typical such configuration is shown in fig. 2. In fig. 2a we plot the monopole loop. The loop consists of 18 links which lie in a two-dimensional plane. We define this plane to be the $(x_1, x_4)$-plane. Also shown is the position of the center of the instanton, which is indicated by the dashed circle. The center is defined to be at the maximum of the action density. In fig. 2b we show the magnetic charge $m(f(x,4))$ and the total electric charge $e^+(x_4)$, as given by eq. (26), of the monopole. We find that the electric charge is indeed equal to $m$, the magnetic charge of the monopole, as we expected. For an anti-instanton we find the electric charge to be $-m$. In fig. 2c we show the electric charge profile of the monopole loop,

$$e^{\pm}(x_4, x_1) = \sum_{x_2, x_3} e^{\pm}(f(x, 4)). \quad (27)$$

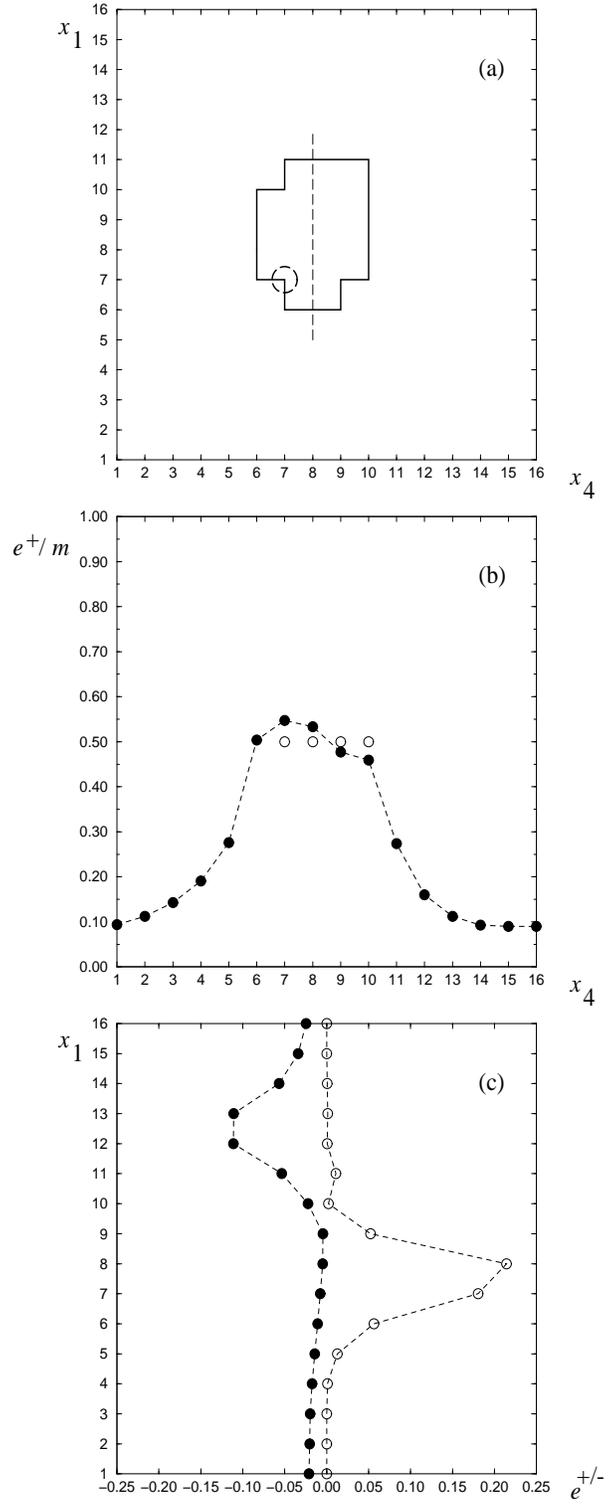

Fig. 2. A single instanton. (a) The monopole loop and position of the instanton (dashed circle). (b) The charge $e^+(x_4)$ (○) and m (●) versus $x_4$. (c) The charge profile $e^\pm(x_4, x_1)$ for $x_4 = 8$.

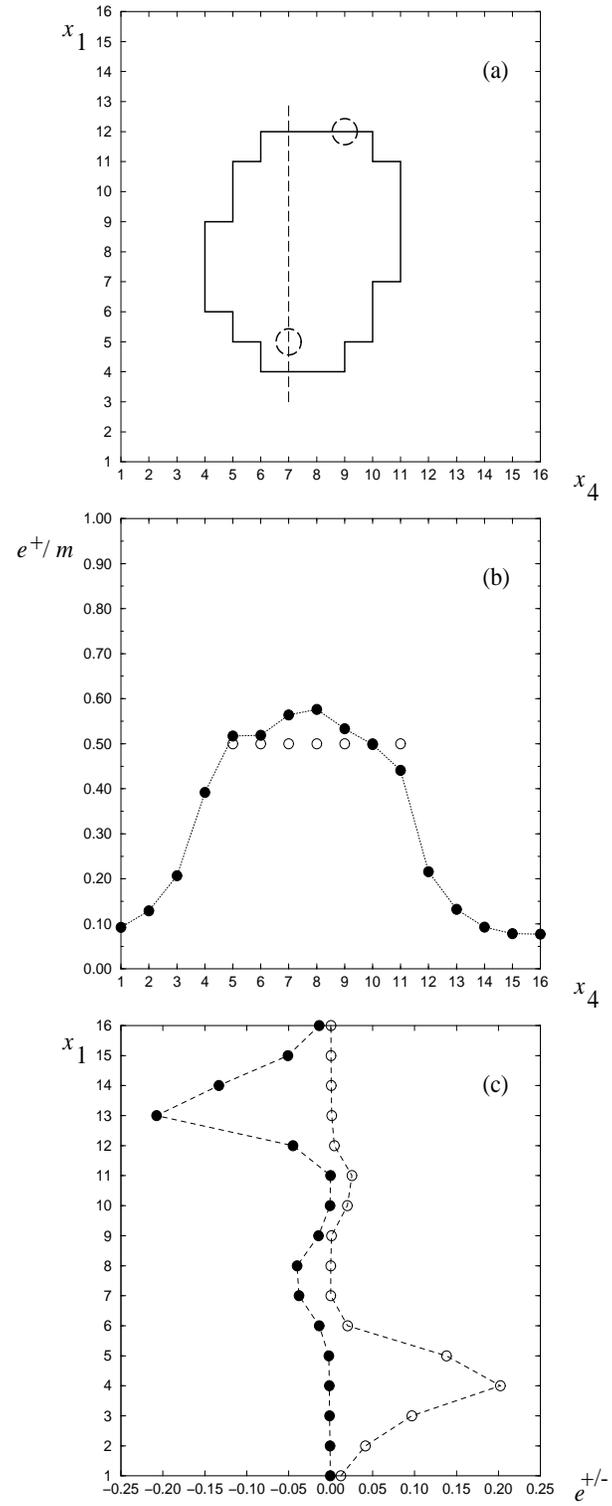

Fig. 3. Two separated instantons. (a) The monopole loop and positions of the instantons (dashed circles). (b) The charge $e^+(x_4)$ (○) and m (●) versus $x_4$. (c) The charge profile $e^{\pm}(x_4, x_1)$ for $x_4 = 7$.

This is the electric charge density summed over two spatial directions, $x_2$ and $x_3$, as a function of $x_1$. We have chosen $x_4 = 8$, which corresponds to the dashed line in fig. 2a. We find that the electric charge is confined to a small region of two to three lattice spacings in diameter about the position of the monopole. The same holds for the charge profile along the other two directions. Thus we may conclude that magnetic and electric currents coincide.

Let us now look at the configuration of two well separated instantons. Such a case is shown in fig. 3. The topological charge of this configuration is $Q = 2$. In fig. 3a we show the monopole loop relative to the centers of the two instantons. Both instantons have the same orientation in group space. We find a single large planar loop encompassing both instantons, rather than two individual loops. As before, we define the plane of the loop to be the $(x_1, x_4)$-plane. The loop consists of 30 links. In fig. 3b we show $m(f(x, 4))$ and $e^+(x_4)$. We find that magnetic and electric charges are equal also in this case. Finally, in fig. 3c we show the electric charge profile $e^{\pm}(x_4, x_1)$ of the monopole loop. Here we have chosen $x_4 = 7$, which is indicated by the dashed line in fig. 3a. As in the single instanton case, the electric charge is localized in a small region about the position of the monopole. We find the same result for the charge profile along the other two directions.

## 3. Discussion

This confirms our expectations: gauge fixing and abelian projection turn instantons into dyons with electric charge $e = \pm m$. It has been shown by Hart and Teper [19] that the size of the monopole loop increases with the (core) size of the instanton. Thus large isolated instantons give rise to large dyon loops. From lattice work [4] one would expect to find at least $O(10)$ (anti-)instantons per $fm^4$. This has interesting consequences.

It was assumed that the dynamical variables of the abelian theory were 'photons', color electrically charged particles and monopoles. By what we know now this list is incomplete. The full list of charges is given in fig. 4 for gauge group $SU(2)$, consisting of 'photons', quarks, gluons and a charge triplet of dyons. Possible bound states are not counted here.

Our examples of large, isolated instantons were textbook constructions. In the confining vacuum we would not expect to find such field configurations, because large loops of electric currents are suppressed exponentially by the area they encompass. For the same reason a dilute instanton gas would hardly be compatible with confinement.

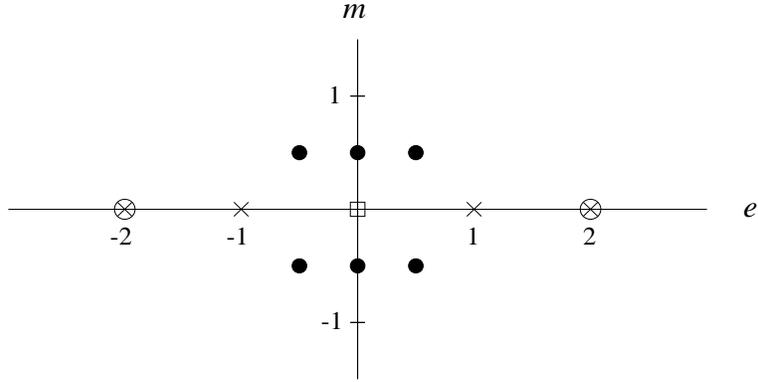

Fig. 4. The electric ($e$) and magnetic ($m$) charge lattice. The symbols denote [16] 'photon' (□), quark (×), gluon (⊗) and dyon (●).

Instead we would expect that the vacuum is a coherent plasma of instantons and anti-instantons in which the dyons can hop from one (anti-)instanton to the next, thereby freely exchanging electric charge so that color electric charge is not transported over macroscopic distances. To make this possible, instantons and anti-instantons must not only strongly overlap, but must also be strongly correlated. In the first place instantons must be surrounded by anti-instantons, and vice versa. Furthermore, adjacent instanton–anti-instanton pairs must have roughly the same orientation in group space. The net effect is sketched in fig. 5 for a single instanton–anti-instanton pair: while the (induced) magnetic charge can move freely, in accord with the dual superconductor picture of confinement, the electric currents are only short-lived. This setup is quite different from a liquid, which by definition has only short-range correlations.

An interesting question is now whether instantons are the only source for monopoles. To find that out, one will have to look for instanton-monopole correlations in the quantum vacuum [27]. Another point of interest is the effective action. The ansätze based on either instantons [8] or monopoles [12,13] should definitely be generalized to include dyons.

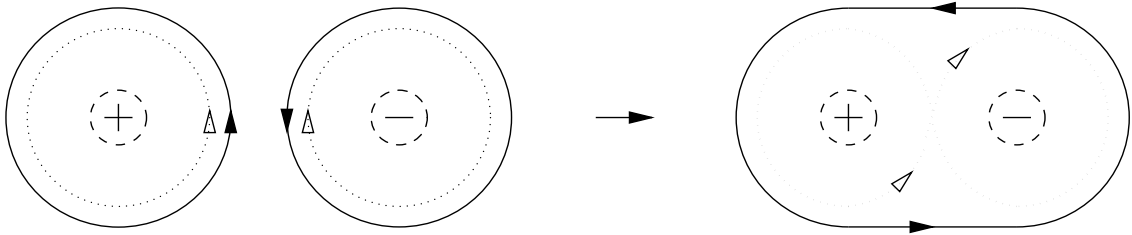

Fig. 5. An instanton and anti-instanton pair (dashed circles) getting attracted. Both point into the same direction in group space. The solid curves denote the magnetic currents, the dotted curves the electric currents.

## Acknowledgments

We like to thank Tsuneo Suzuki and his colleagues for their kind hospitality at Kanazawa University where this paper was completed. The work of VB was supported in part by grant Nos. NJP100 and NJP300 of the International Science Foundation and by grant No. 96-02-17230-a of the Russian Foundation for Fundamental Sciences.